\documentclass[10pt]{article}
\usepackage{amsfonts}
\usepackage{amssymb}
\usepackage{amsthm}

\newcommand{\sss}{\setcounter{equation}{0}}
\newtheorem{theorem}{THEOREM}[section]

\newtheorem{lemma}[theorem]{LEMMA}

\newcommand{\ere}{ {\mathbb R}}

\newcommand{\ese}{{\mathbb S}}
\newcommand{\CE}{{\mathbb C}}
\newcommand{\ls}{L^2(\ese^{n-1})}
\def\beq{\begin{equation}}
\def\ene{\end{equation}}

\def \ds {\displaystyle}
\newcommand{\bull}{\hfill $\Box$}

\def\qed{\ifhmode\unskip\nobreak\fi\ifmmode\ifinner
\else\hskip5pt\fi\fi\hbox{\hskip5pt\vrule width4pt height6pt
depth1.5pt\hskip1pt}}

\def\sc0{s_0(\omega,\omega^{\prime}; \lambda)}

\begin{document}
\baselineskip=20 pt
\parskip 6 pt

\title{Completeness of Averaged Scattering Solutions and Inverse
Scattering at a Fixed Energy
 \thanks{ Research partially supported by
Universidad Nacional Aut\'onoma de M\'exico under Project
PAPIIT-DGAPA IN 105799, and by CONACYT under Project P42553­F.}}
 \author{ Ricardo Weder\thanks{ †Fellow Sistema Nacional de Investigadores.}
\\Instituto de Investigaciones en Matem\'aticas Aplicadas y en Sistemas \\Universidad Nacional Aut\'onoma de
M\'exico \\
Apartado Postal 20-726, M\'exico DF 01000
\\weder@servidor.unam.mx}

\date{}
\maketitle
\begin{center}
\begin{minipage}{5.75in}
\centerline{{\bf Abstract}}
\bigskip

 We prove that the averaged scattering solutions to the
Schr\"odinger equation with short-range electromagnetic potentials
$(V,A)$ where $V(x)=O(|x|^{-\rho}), A(x)= O(|x|^{-\rho}), |x|
\rightarrow \infty, \rho >1,$ are dense in the set of all
solutions to the Schr\"odinger equation that are in $L^2(K)$ where
$K$ is any connected bounded open set in $\ere^n,n\geq 2,$ with
smooth boundary.

We use this result to prove that if two short-range
electromagnetic potentials $(V_1,A_1)$ and $(V_2,A_2)$ in $\ere^n,
n\geq 3,$ have the same scattering matrix at a fixed positive
energy and if the electric potentials $V_j$ and the  magnetic
fields $ F_j:={\rm curl} A_j , j=1,2,$ coincide outside of some
ball they necessarily coincide everywhere.

In a previous paper of Weder and Yafaev the case of electric potentials and magnetic fields that are asymptotic
sums of homogeneous terms at infinity was studied. It was proven that all these terms can be uniquely reconstructed
from the singularities  in the forward direction of the scattering amplitude at a
fixed positive energy.

The combination of the new uniqueness result of this paper and the
result of Weder and Yafaev implies that the scattering matrix at a
fixed positive energy uniquely determines electric potentials and
magnetic fields that  are a finite sum of homogeneous terms at
infinity, or more  generally, that are asymptotic sums of
homogeneous terms that actually converge, respectively, to the
electric potential and to the magnetic field.

\end{minipage}
\end{center}
\newpage

\section{Introduction}\sss
Let us first consider the stationary Schr\"odinger equation

\beq
-\Delta \phi + V(x)\phi = E \phi, E=k^2, k >0,
\label{1.1}
\ene
on $\ere^n, n \geq 2,$ where the real-valued potential $V$ satisfies,

\beq
V(x)=O(|x|^{-\rho}), |x|\rightarrow \infty, \rho >1.
\label{1.2}
\ene
Later we give precise conditions on $V$ and we introduce magnetic potentials.

If $\rho > n,$ for any $E>0$ and any unit vector $\omega \in
\ese^{n-1}$ (\ref{1.1}) has a unique solution $\phi_+(x,\omega;E)$
with the asymptotics as $|x|\rightarrow \infty$ \beq
\phi_+(x,\omega;E)=e^{ik x\cdot\omega}+f
\frac{e^{ik|x|}}{\ds|x|^{(n-1)/2} }+o(|x|^{-(n-1)/2}). \label{1.3}
\ene If we only have that $\rho > (n+1)/2$ the unique solution
with the asymptotics (\ref{1.3}) exists but now the
$o(|x|^{(n-1)/2})$ has to be interpreted in an appropriate
averaged sense. For a discussion of this issue see, for example,
\cite{yaf1}. The unique solution with the asymptotics (\ref{1.3})
is called the scattering solution. The coefficient
$f=f(\nu,\omega;E)$ depends upon the incident direction $\omega$
of the incoming plane wave $e^{ik x\cdot\omega}$ its energy,
$E=k^2,$ and the direction $\nu:= x/|x|$ of observation of the
outgoing spherical wave $e^{ ik|x|}\big/ |x|^{(n-1)/2} $.
The function $f(\nu,\omega;E)$ is known as the scattering
amplitude. In potential scattering in quantum mechanics the plane
wave describes a beam of particles incident on a scattering center
described by $V$, and the outgoing spherical wave corresponds to
the scattered particles.

The unitary operator that corresponds to this scattering process is the scattering matrix, that is the
unitary operator in $L^2(\ese^{n-1})$ that is defined in terms of the scattering amplitude by the formula,

\beq
( S( E) u )(\omega)=u (\omega)+ i e^{i
 \pi(n-3)/4}E^{(n-1)/4} \left(2\pi\right)^{-(n-1)/2}\int_{\ese^{n-1}}
 f(\nu,\omega; \lambda) u(\omega) \, d\omega.
\label{1.4} \ene
In the case of general short-range potentials that satisfy (\ref{1.2}) with $\rho >1$ the scattering amplitude
can be defined by (\ref{1.4}) where the scattering matrix  $S(E)$ is defined via the time-dependent wave operators.
See Section 2.

An important property of the scattering solutions is that for any connected bounded open
set $K \subset \ere^n, n\geq 2,$ with smooth boundary, the set

\beq \left\{ \phi_+(x,\omega;E)  \right\}_{\ds \omega \in
\ese^{n-1}} \label{1.5} \ene
is strongly dense in the set of all
solutions to (\ref{1.1}) that are in $L^2(K)$. This was proven by
D. Eidus in \cite{ei} for bounded potentials that satisfy
(\ref{1.2}) with $\rho >n$ and, independently,  \cite{we1} proved a
similar result. The proof in \cite{we1} applies to potentials that
satisfy (\ref{1.2}) with $ \rho > (n+1)/2$. The density of the
scattering solutions was applied in \cite{ei} to prove that
(\ref{1.1}) has the Runge property and in \cite{we1} to prove that
 if two electric potentials
in $\ere^n, n\geq 3,$ have the same scattering matrix at a fixed
positive energy and if they coincide outside of some ball they
necessarily coincide everywhere. For further references on the
completeness of the scattering solutions and for the application
to inverse  boundary value and inverse scattering problems see
\cite{su}, \cite{ra}, \cite{is} and the references quoted there.

In this paper we wish to generalize the completeness of the scattering solutions to the
case of potentials that satisfy (\ref{1.2}) with $\rho >1$. The first problem that we have
to address is that in general (\ref{1.1}) has no solutions with the asymptotics (\ref{1.3}) if
(\ref{1.2}) only holds for some $\rho >1$. For a discussion of this issue see  \cite{skr} and \cite{yaf1}.
To see what would be an appropriate generalization we observe that the completeness of the scattering solutions
is equivalent to the completeness of the set of solutions

\beq
\phi_{+,f}(x;E):= \int_{\ds \ese^{n-1}} \phi_+(x,\omega;E)\, f(\omega)\, d\omega, f \in L^2(\ese^{n-1})
\label{1.6}
\ene
that are obtained by taking the average on the angular variables of the scattering solutions
with arbitrary functions in $L^2(\ese^{n-1})$. The point is that in the general case where (\ref{1.2}) only holds
for some $\rho >1$ it is possible to define solutions to (\ref{1.1}) $\phi_{+,f}(x;E)$ for all $f \in L^2(\ese^{n-1})$
that for regular $f$ are asymptotic to a linear combination of incoming and outgoing spherical waves and furthermore,
the action of the scattering matrix $S(E)$ can be determined in terms of these solutions. Moreover, if (\ref{1.2})
holds with $\rho > (n+1)/2$ the solutions $\phi_{+,f}$ are given by the right-hand side of (\ref{1.6}). For these results
see \cite{yaf2} -where also a generalized eigenfunctions expansion theorem in terms of these solutions is proven- and
\cite{ag}.

Related problems appear in different settings. See \cite{we2} for
the case of acoustic and electromagnetic waves in perturbed
stratified media.

We prove  in Theorem 3.1  the completeness of the averaged
scattering solutions  $\phi_{+,f}$ in the case of the stationary
Schr\"odinger equation with electric potential $V$ and magnetic
potential $A$,
\beq\mathrm{}
\left( i\nabla+A \right)^2 \phi+V\phi=E \phi, E=k^2, k>0
\label{1.7}
\ene
where $V$ satisfies (\ref{1.2}) and

\beq A(x)=O(|x|^{-\rho}), \partial_j A^{(l)} =O(|x|^{-\rho}),
|x|\rightarrow \infty,1\leq j,l \leq n,
 \label{1.8}
 \ene
 for some $\rho >1$. For
precise conditions see Theorem 3.1.

With the help of our result on the completeness of the averaged scattering solutions we prove in Theorems 4.2 and 4.3
 that if two electromagnetic potentials $(V_1,A_1)$ and  $(V_2,A_2)$ in $\ere^n, n\geq 3,$ that satisfy
(\ref{1.2}) and (\ref{1.8}) with $\rho >1$ have the same
scattering matrix at a fixed positive energy and if the electric
potentials $V_j$ and the the magnetic fields $ F_j:={\rm curl} A_j
, j=1,2,$ are equal outside of some ball, then $V_1= V_2$ and
$F_1= F_2$ everywhere. This generalizes our result of \cite{we1}
in two directions. First, we allow now for general short-range
electric potentials that satisfy (\ref{1.2}) with $\rho >1$ and
second, we consider now the case where there is also a short-range
magnetic potential that satisfies (\ref{1.8}) with $\rho >1$. This
decay is optimal for short-range potentials. Note that as the
scattering matrix is invariant under short-range gauge
transformations uniqueness of $A$ does not hold

We recall that in  Theorem 4.2 of \cite{wy} the case of electric potentials and magnetic fields that are asymptotic
sums of homogeneous terms at infinity was studied. It was proven that all these terms can be uniquely reconstructed
from the singularities  in the forward direction of the scattering amplitude at a
fixed positive energy.

By combining our new uniqueness result in Theorem 4.3  with
Theorem 4.2 of \cite{wy} we prove in Theorem 4.4 that the
scattering matrix at a fixed positive energy uniquely determines
electric potentials and magnetic fields in $\ere^n, n\geq 3,$ that
are  finite sums of homogeneous terms at infinity, or more
generally, that are asymptotic sums of homogeneous terms at
infinity that actually converge, respectively, to the electric
potential and to the magnetic field. This result generalizes
Theorem 4.6 of \cite{wy} to the case where there is also a
magnetic potential and where both the electric potential and the
magnetic field satisfy the optimal short-range decay condition.

It is known since quite some time that the scattering matrix at a fixed positive energy uniquely determines
electric potentials and magnetic fields if strong restrictions on the decay at infinity are imposed.
The paper \cite{nus} considers potentials of
compact support, and  \cite{no, er, iso, uh}  potentials
decaying exponentially at infinity. On the contrary, for general
short-range potentials the scattering matrix at a fixed positive
energy does not determine uniquely the potential. Indeed, in
\cite{cs} examples -in three dimensions- are given of non-trivial
radial oscillating potentials with  decay as $|x|^{-3/2}$ at
infinity such that the corresponding scattering amplitude is identically zero at some positive energy. Moreover,
in dimension two there are examples \cite{Gr} of potentials with a regular decay as $|x|^{-2}$ at infinity that
have zero scattering amplitude at some positive energy. Nevertheless, as we discussed above if two  general
short-range electric potentials and magnetic fields coincide outside of some ball and if they have the same scattering
matrix at some positive energy they are equal everywhere.

Actually, the same problem appears in different settings. Thus, it is proven in \cite{iso2, we3, gr}
that the scattering matrix at a fixed positive energy uniquely
determines an exponentially decreasing perturbation of a stratified
media. As another example, we mention that the scattering matrix at a fixed quasi-energy uniquely
determines time-periodic potentials that decay exponentially at spatial
infinity \cite{we4}.

Theorem 4.6 of \cite{wy} and its generalization in Theorem 4.4 below show a new aspect of the inverse scattering problem
at a fixed energy. Namely, that uniqueness holds for general short-range electric potentials and magnetic fields without
strongly restricting the decay at infinity, provided that the electric potential and the magnetic field have a regular
behaviour at infinity. Of course, this eliminates the oscillations and hence there is no contradiction with the examples
of \cite{cs}. Furthermore, as we consider three or more dimensions there is no contradiction with the two dimensional
examples of \cite{gr}.

The paper is organized as follows. In Section 2 we discuss some basic results on the limiting absorption principle
and on stationary scattering theory, we consider the averaged scattering solutions and we give a representation of the
scattering matrix in terms of these solutions. In Section 3 we prove Theorem 3.1 on the completeness of the averaged
scattering solutions by generalizing the proof given in \cite{we1}. In Section 4 we prove Theorems 4.2 and 4.3
extending to this case the proof of Theorem 1 of \cite{we1} and, finally, we prove  Theorem 4.4.

\section{Basic Results}
\sss In this section we recall some well known results on the
stationary scattering theory of the Schr\"odinger operator  in
$\ere^n,n\geq2,$ with short-range electromagnetic potentials
\cite{kur,ag,rs,kur2,yaf4}. For any $\alpha \in \ere$ let us denote by $\mathbb H^{\alpha}= \mathbb H^{\alpha,2}$ the
$L^2$-based Sobolev space. See
for example \cite{sch}.

We consider the Schr\"odinger operator,

\beq H:= \left( i\nabla+A \right)^2+V = H_0 + Q, \
\label{2.1}
\ene

where the free Hamiltonian, $H_0:=-\Delta$ is a self-adjoint
operator with domain the Sobolev space $\mathbb H^2$ and

\beq
Q:= 2i A\cdot \nabla +i{\rm Div}A + A^2 +V
\label{2.2}
\ene
is the perturbation.

In this section we always assume that for some $\epsilon >0$ the operator $(1+|x|)^{1+\epsilon}Q$
is compact from  $\mathbb H^2$ into $L^2$. Sufficient conditions that assure that this is true are well known.
See for example \cite{sch}. In particular, it follows from Theorem 5.2 of \cite{sch} that this is the case if the
following is true. If $n=2,3, V, {\rm Div A},A^2 \in L^2_{\rm loc},$  if $n=4, V, {\rm Div}A,
A^2 \in L^{2+\delta}_{\rm loc}$ for some $\delta >0$ and if $ n\geq 5, V, A,{\rm Div A}, A^2 \in L^{n/2}_{\rm loc}$ and if
moreover, for some constants $C, R >0 $,

\beq
|V(x)|+|A(x)|+|\partial_j A^{(l)}(x)| \leq C (1+|x|)^{-\rho},
\,\rho
>1, 1\leq j,l\leq n, \,{\rm for}\, |x| \geq R > 0. \label{2.3} \ene

Under these conditions the Schr\"odinger operator $H$ is self-adjoint and bounded below with domain $\mathbb H^2$. It has no singular
continuous spectrum and its absolutely-continuous spectrum is $[0,\infty)$. By unique continuation \cite{ho}, \cite{wo},
and Theorem 1.2 of \cite{au} $H$ has no positive eigenvalues. The negative spectrum consists of
eigenvalues with finite multiplicity and they can only accumulate at zero.

To state the limiting absorption
principle we introduce weighted $L^2$ spaces for $s \in \ere$.

$$
L^2_s:=\left\{ f: (1+|x|^2)^{s/2} f(x)\in L^2\right\}, \|f\|_{\ds L^2_s}:= \|   (1+|x|^2)^{s/2} f(x )\|_{\ds L^2},
$$
and for any $ \alpha, s \in \ere$,

$$
\mathbb H^{\alpha,s}:= \left\{ f(x): (1+|x|^2)^{s/2} f(x)\in \mathbb H^{\alpha}\right\},
\|f\|_{\ds \mathbb H^{\alpha,s}}:= \| (1+|x|^2)^{s/2} f(x)\|_{\ds \mathbb H^{\alpha}}.
$$
 $\CE^{\pm}$ denotes, respectively, the upper, lower,
 complex half-plane.

 The limiting absorption principle is the
following statement. For $z$ in the resolvent set of $H$ let $R(z):= (H-z)^{-1}$ be the resolvent. Then, for every
$E \in (0,\infty)$ the following limits,

$$
R(E\pm i0):= \lim_{\epsilon \downarrow 0}R(E\pm i\epsilon),
$$
exist in the uniform operator topology in $\mathcal B\left( L^2_s, \mathbb H^{\alpha,-s} \right), s >1/2, \alpha \leq 2$,
where for any pair of Banach spaces $X,Y,\, \mathcal B(X,Y)$ denotes the Banach space of all bounded operators
 from $X$ into $Y$.
The functions,

$$
R_{\pm} (E) := \cases{ R(E),    & ${\rm Im}\,
E \not= 0 $, \cr\cr R(E\pm i0) & , $ E\in (0,\infty)
$,}
$$
defined for $E \in \CE^{\pm}\cup (0,\infty)$  with values in $\mathcal B \left(L^2_s,
\mathbb H^{\alpha,-s}\right)$ are analytic for ${\rm Im} \,E \not = 0$ and locally H\"older continuous for
$ E\in (0,\infty)$ with exponent $\gamma$ satisfying $\gamma < 1, \gamma < s-1/2$.

The wave operators,

$$
W_{\pm}:= {\rm s}-\lim_{t\rightarrow \pm \infty} e^{itH}\, e^{-it H_0}
$$
exist as strong limits and are complete, i.e., Range
$W_{\pm}=\mathcal H_{ac}$ where $\mathcal H_{ac}$ denotes the
subspace of absolute continuity of $H$. Moreover, they have the
intertwining property, $H W_{\pm}=W_{\pm} H_0$. The scattering
operator,

$$
\mathbf S:= W_+^{\ast}\, W_-
$$
is unitary.

Let us denote by $T_0(E)$ the following trace operator,

\beq (T_0(E)\phi)(\omega):= 2^{-1/2}\,E^{\ds(n-2)/4}
\frac{1}{(2\pi)^{n/2}} \int_{\ere^n} e^{\ds -iE^{1/2 }
x\cdot\omega} \, \phi(x) \, dx,
\label{2.4}
\ene
that is bounded from $L^2_s, s > 1/2,$ into $L^2(\ese^{n-1})$, and furthermore, the operator valued
function $E \rightarrow T_0(E)$ from $(0,\infty)$ into $\mathcal B (L^2_s,L^2(\ese^{n-1}))$ is locally
H\"older continuous with exponent $\gamma < 1, \gamma < s-1/2$.
Moreover, the operator,

\beq
\left(\mathcal F_0 \phi\right)(E, \omega):= \left(T_0(E)\phi\right)(\omega),
\label{2.5}
\ene
extends to a unitary operator from $L^2$ onto $\hat{\mathcal H}:= L^2((0,\infty);L^2(\ese^{n-1}))$ that
gives a spectral representation for $H_0$, i.e.,

\beq
\mathcal F_0 H_0 \mathcal F_0^{\ast}= E,
\label{2.6}
\ene
the operator of multiplication by $E$ in $\hat{\mathcal H}$.

The perturbed trace operators are defined as follows,

\beq
\left(T_{\pm}(E)\phi\right)(\omega):= T_0(E)(I-QR_{\pm}(E))\phi,
\label{2.7}
\ene
for $E \in (0,\infty)$. They are bounded from $L^2_s, s > 1/2,$ into $L^2(\ese^{n-1})$, and furthermore,
the operator valued functions $E \rightarrow T_{\pm}(E)$ from $(0,\infty)$ into
$\mathcal B (L^2_s,L^2(\ese^{n-1}))$ are locally
H\"older continuous with exponent $\gamma < 1, \gamma < s-1/2$.
The operators,
\beq
\left(\mathcal F_{\pm} \phi\right)(E, \omega):= \left(T_{\pm}(E)\phi\right)(\omega),
\label{2.8}
\ene
extend to  unitary operators from $\mathcal H_{ac}$ onto $\hat{\mathcal H}$ and they give
 spectral representations for the restriction of $H$ to $\mathcal H_{ac}$,

\beq
\mathcal F_{\pm} H \mathcal F_{\pm}^{\ast}= E
\label{2.9}
\ene
the operator of multiplication by $E$ in $\hat{\mathcal H}$. Furthermore, the stationary formulae for
the wave operators hold,

$$
W_{\pm}= \mathcal F_{\pm}^{\ast}\, \mathcal F_0.
$$
As $\mathbf S$ commutes with $H_0$ we have that,
$$
\left(\mathcal F_0 S \mathcal F_0^{\ast} \phi\right)(E,\omega)= S(E)\phi,
$$
where   $S(E), E>0,$ is unitary on $L^2(\ese^{n-1})$. The operator $S(E)$ is the scattering matrix. This time-dependent
definition of the scattering matrix generalizes to general short-range potentials the definition given in Section 1.

The scattering matrix has the following stationary representation,

\beq S(E)= I-2\pi i \mathcal F_0 \, Q \left[I- R_+(E)\, Q\right]\,
\mathcal F_0^{\ast}, E\in (0,\infty). \label{2.10}
\ene

The scattering matrix can be represented in terms of averaged scattering solutions as follows (see \cite{we2}
for a similar representation in the case of acoustic and electromagnetic waves in perturbed stratified media).

For any $ f \in \ls$ let us define the unperturbed averaged scattering solutions as follows,
\beq
\phi_{0,f}(x;E):= \int_{\ds \ese^{n-1}}
e^{i E^{1/2}x\cdot \omega}
\, f(\omega)\, d\omega.
\label{2.11}
\ene

Observe that $ \phi_{0,f} \in L^2_{-s}, s > 1/2,$ and that  $H_0 \phi_{0,f}= E \phi_{0,f}$.
The perturbed averaged scattering solutions are defined as,

\beq
\phi_{+,f}(x;E):= [I- R_+(E)Q]\phi_{0,f}, \, E \in (0,\infty), \, f \in L^2(\ese^{n-1}).
\label{2.12}
\ene

Then, $ \phi_{+,f} \in L^2_{-s}, s > 1/2,$ and   $H \phi_{+,f}= E \phi_{+,f}$.

By (\ref{2.10}) for $f,g \in \ls$,

\beq
\left(S(E)f,g \right)_{\ds\ls}= (f,g)_{\ds\ls}- i \frac{\ds E^{(n-2)/2}}{\ds2 (2\pi)^{n-1}}
\left(Q \, \phi_{+,f}, \phi_{0,g}\right)_{\ds L^2}.
\label{2.13}
\ene

If $\rho > (n+1)/2$ in (\ref{2.3}), $V, A,A^2,{\rm Div}A \in
L^2_s, s >1/2,$ and we can define the scattering solution,

$$
\phi_+(x,\omega;E):= e^{iE^{1/2}x\cdot \omega}-R_+(E)\left( Q e^{iE^{1/2}x\cdot \omega}  \right).
$$
In this case
$$
\phi_{+,f}(x;E)= \int_{\ese^{n-1}} \phi_+(x,\omega;E)\, f(\omega)\, d\omega,
$$
what justifies the name averaged scattering solutions. See \cite{ag} for further discussions on this point.

\section{Completeness of Solutions}
\sss In this section we prove our result on the completeness of
the averaged scattering solutions.

\begin{theorem}\label{t.2.1}
Suppose that if  $n=2,3, V, {\rm Div A},A^2 \in L^2_{\rm loc},$  if $n=4, V, {\rm Div}A,
A^2 \in L^{2+\delta}_{\rm loc}$ for some $\delta >0$ and if $ n\geq 5, V, A,{\rm Div A}, A^2 \in L^{n/2}_{\rm loc}$
and that

\beq
|V(x)|+|A(x)|+|\partial_j A^{(l)}(x)| \leq C (1+|x|)^{-\rho}, \,\rho >1, 1 \leq j,l \leq n,\,{\rm for}\, |x| \geq R > 0.
\label{3.0}
\ene
Let $K$ be a
connected open bounded set with smooth boundary. Then, the set of averaged scattering solutions,
$\phi_{+,f}, f \in \ls$,  is strongly dense on the set of all
solutions to (\ref{1.7}) in $L^2(K)$.
\end{theorem}
\noindent{\it Proof:}
We follow the proof given in \cite{we1}. Suppose that $ \phi\in L^2(K)$ is a solution to  (\ref{1.7})
that is orthogonal to all the averaged scattering solutions, i.e.,
\beq
\left( \varphi, \phi_{+,f}\right)_{\ds L^2(K)}=0, f \in \ls,
\label{3.1}
\ene
and define,
\beq
\psi:= R_+(E)\varphi,
\label{3.2}
\ene
where we have extended $\varphi$ by zero to $\ere^n\setminus K$.

By (\ref{2.9}), and as the trace operator $T_-(E)$ is locally
H\"older continuous, it follows from  Privalov's theorem that,
\beq
\psi=\psi_1+\psi_2,
\label{3.3}
\ene
where,

\beq \psi_1:= {\rm P.V.}\int_{I}\, d\lambda \, \frac{1}{\lambda
-E}\,T_-^{\ast}(\lambda)T_-(\lambda)\varphi + i\pi
T_-{^\ast}(E)T_-(E)\varphi, \label{3.4} \ene \beq \psi_2 = R(E) \,
{\mathcal E}(\tilde{I})\varphi, \label{3.5} \ene

where $I:=[a,b], 0 < a < E-\delta, b > E+\delta$ for some $\delta >0, \tilde{I}:= \ere\setminus I,$ and
${\mathcal E}(\cdot)$ is the spectral family of $H$.

Clearly,
\beq
\psi_2 \in L^2.
\label{3.6}
\ene
Moreover, by (\ref{2.7}) and (\ref{2.12}) equation (\ref{3.1}) implies that,
\beq
T_-(E)\varphi=0,
\label{3.7}
\ene
and  as $T_-$ is H\"older continuous we can eliminate the P.V. in the integral in the right-hand side of (\ref{3.4})
and then,

\beq \psi_1=\int_{I}\, d\lambda \, \frac{1}{\lambda
-E}\,T_-^{\ast}(\lambda)T_-(\lambda)\varphi.
\label{3.8}
\ene
Let us denote by $J_I$ the operator,
\beq J_I:= \mathcal E(I)\,
\mathcal F_-^{\ast}= \mathcal F_-^{\ast}\, \chi_I(E),
\label{3.9}
\ene
where $\chi_I$ is the characteristic function of $I$. Since
$\mathcal F_-$ is unitary from $\mathcal H_{ac}$ onto
$\hat{\mathcal H}$ and (\ref{2.9}) holds, it follows that $J_I$ is
unitary from $L^2(I, \ls)$ onto $\mathcal E(I)\mathcal H_{ac},$
and in consequence it is bounded from $L^2(I, \ls)$ into $L^2$.
Moreover, by (\ref{2.8})
\beq J_I \varphi= \int_I\,
T_-^{\ast}(\lambda)\, \varphi(\lambda)\, d\lambda,
\label{3.10}
\ene
and it follows that $J_I$ is bounded from  $L^1(I, \ls)$ into
$L^2_{-s},$ for any $ s > 1/2$. Then, by interpolation \cite{rs2}
$J_I$ is bounded from $L^p(I, \ls)$ into $L^2_{\ds -\epsilon_p s},
\epsilon_p:= \frac{2}{p}-1, 1\leq p\leq 2$. Observe that by
(\ref{3.7}), (\ref{3.8}) \beq \psi_1= J_I \frac{1}{\lambda
-E}\,\left( T_-(\lambda)- T_-(E)\right) \varphi. \label{3.11} \ene

Let us take $ s_0:=1/2+ \gamma$, where $ \gamma < {\rm min} [1/2,
\rho/2]$. Since $T_-$ is locally H\"older continuous from
$L^2_{ s_0}$ into $\ls$ with exponent $ \gamma$, and as
$\varphi \in L^2_{ s_0}$ it follows  that,

$$
\frac{1}{\lambda -E}\,
\left(T_-(\lambda)-T_-(E)\right)\varphi \in  L^p(I, \ls), p < 1/(1-\gamma)
$$
and by taking $s$ close enough to $1/2$ we conclude that $\psi_1 \in L^2_{-\beta}$ for some $ 0< \beta < 1/2$.
Hence, by (\ref{3.3}) and (\ref{3.6}) $\psi \in L^2_{-\beta}$. Furthermore,

$$
H \psi = E \psi + \varphi,
$$
and as $\varphi(x) =0,$ for $x \in \ere^n\setminus K$ it follows
from Theorem 1.2 of \cite{au} that $\psi (x)$ is identically zero
in the complement of a large enough ball and then, by unique continuation
 \cite{ho}, \cite{wo} it is identically zero on $\ere^n \setminus K$. In
particular, $\psi(x)= \nabla \psi(x)=0$ on $\partial K$ in trace
sense. Finally, approximating $\psi$ in the norm of $\mathcal
H^2(K)$ by functions in $C^{\infty}_0(K$) (it is here that the
smoothness of $\partial K$ is used) we prove that,

$$
\|\varphi\|^2_{\ds L^2(K)}= \left((H-E)\psi,\varphi\right)_{\ds L^2(K)}=
\left(\psi, (H-E)\varphi\right)_{\ds L^2(K)}=0,
$$
and it follows that $\varphi =0$.

\section{Inverse Problem }
\sss Note that it follows from the definition of the wave
operators that the scattering operator $\mathbf S$ and the
scattering matrix  $S(E)$ are invariant under the gauge
transformation, $A \rightarrow A+ \nabla \psi$, where $|\psi(x)|
\leq C(1+|x|)^{-\mu}, |\nabla \psi (x)| \leq C(1+|x|)^{-1-\mu}, \mu
> 0$. This invariance suggest that we should associate the
scattering operator and the scattering matrix to the magnetic
field $F={\rm curl}A$. The problem is that in the Schr\"odinger
equation  (\ref{1.7}) as well as in the definition of the
Hamiltonian (\ref{2.1}), the magnetic potential appears
explicitly and, in general, it is not possible to express $\mathbf
S$ and $S(E)$ only in terms of $F$. A striking manifestation of
this fact is the famous Aharonov-Bohm effect \cite{ab}. For a
study of inverse scattering in the context of the Aharonov-Bohm
effect see \cite{we5}.

In our case we can proceed as follows.
We consider $\ere^n, n\geq 3$, and we assume  that $A$ satisfies,

\beq \left|\partial^{\alpha} A(x)\right| \leq C_{\alpha}
(1+|x|)^{-\rho-|\alpha|},\quad \rho >1, 0 \leq |\alpha| \leq 1.
\label{4.1}
\ene

We use a three-dimensional notation for the curl and the divergence keeping in mind that in the
general case $A$ is a $1$-form and $F$ is a $2$-form.
 By definition, $F(x)= {\rm curl}\, A(x)$, and in  terms of components it is given by,
 \beq
F^{(ij)}(x)= \partial_{ i}A^{(j)}(x)-\partial_{ j}A^{(i)}(x).
 \label{4.2}
 \ene

Note that, ${\rm div} F=0$. Clearly, from a magnetic field $F(x)$ such that
 $ {\rm div}\, F(x)=0$ we can only reconstruct the magnetic potential up to arbitrary gauge transformations.

We find it convenient use the procedure given in  in \cite{yaf3}, \cite{wy} to construct a short-range magnetic potential
for an arbitrary magnetic field satisfying  ${\rm div}F=0$ and the estimate,
 \beq
 \left|\partial^{\alpha} F(x)\right|  \leq C
 (1+|x|)^{- 1-\rho-|\alpha|}, \quad \rho > 1, 0 \leq |\alpha| \leq 1.
\label{4.3}
\ene
 Let us define the auxiliary potentials

\beq
A_{reg}^{(i)} (x) = \int_1^{\infty}s \sum_{j=1}^d
F^{(ij)}(sx)x_j\, ds,
\quad A_{\infty}^{(i)} (x) =
-\int_0^{\infty}s \sum_{j=1}^d F^{(ij)}(sx)x_j\, ds.
\label{4.4}
\ene

Observe that $A_{\infty}$ is a homogeneous function
of order $-1$, and that ${\rm curl}\, A_\infty(x)=0$ for $x\neq 0$. Let us now  define the function $U(x)$
for $x\neq 0$ as a curvilinear integral
\beq
U(x)=\int_{\Gamma_{x_{0},x}} A_{\infty}(y)\cdot dy
\label{4.5}
\ene
taken between some fixed point  $x_{0} \neq 0$ and
a variable point $x$. We require that $0\not\in
\Gamma_{x_{0},x}$. Then, by Stokes theorem,
 the function $U(x)$ does not depend on the choice of the
contour $\Gamma_{x_{0},x}$ and $\nabla U(x)=A_\infty(x)$.

Finally, we  choose an arbitrary function $\eta \in C^\infty
(\ere^n)$ such that $\eta (x)=0$ in a neighbourhood of zero, $\eta
(x)=1$ for $|x|\geq R$, for some $R > 0$, and define,
 \beq
A(x) := A_{reg}(x) + (1- \eta (x)) A_{\infty} (x)-U(x)\nabla \eta (x) .
\label{4.6}
\ene
Then, $   {\rm curl}\, A(x)=F(x)$, $A$ satisfies (\ref{4.1}) and
$A(x) = A_{reg}(x)$ for $|x|\geq R$.

In this section we always associate  to a magnetic field $F$ satisfying ${\rm div} F=0$ and
(\ref{4.3}) the magnetic potential $A$ given by  formulae (\ref{4.4}) --  (\ref{4.6}) and then
construct the scattering operator ${\mathbf S}$ and the  scattering matrix $S(E)$ in terms of the
Schr\"odinger operator (\ref{2.1}) with this potential. If another short-range
potential $\tilde{A}$ satisfies (\ref{4.1}) and moreover, $ {\rm curl}\,
\tilde{A}(x)=F(x)$, then necessarily $A$ and $\tilde{A}$ are related by a gauge transformation and the scattering
operators and scattering matrices corresponding to $A$ and $\tilde{A}$ coincide. It is in this sense that we
 speak about the scattering operator $\mathbf S$ and the scattering matrix $S(E)$
corresponding   to the magnetic field $F$. The key issue  that makes the magnetic potential (\ref{4.4}-\ref{4.6})
important for us is that if two magnetic fields that satisfy (\ref{4.3}) coincide outside of a ball of radius
bigger or equal to $R$, then,
the corresponding magnetic potentials given by (\ref{4.4}-\ref{4.6}) also coincide outside of the same ball.

\begin{lemma}
Let $V_j,j=1,2$ be electric potentials in $\ere^n,n\geq 3,$ such that, if  $n=3, V_j \in L^2_{\rm loc},$
if $n=4, V_j \in L^{2+\delta}_{\rm loc}$ for some $\delta >0$ and if $ n\geq 5, V_j \in L^{n/2}_{\rm loc}$  and
 for some $ R>0$,
\beq
\left| V_j(x) \right|\leq C (1+|x|)^{-\rho}, \quad \rho >1, \,|x| \geq R > 0, j=1,2.
\label{4.7}
\ene
Furthermore, let $F_j, j=1,2,$ be magnetic potentials that satisfy (\ref{4.3}). Let $S_j(E)$ be
the scattering matrices corresponding, respectively, to $(V_j,F_j), j=1,2$.
Suppose that for some $E >0$, $S_1(E)= S_2(E)$ and that for some $R_1 >0, V_1(x)=V_2(x), F_1(x)=F_2(x)$ for
$|x|\geq R_1 >0$.
Then, the averaged scattering solutions $\phi_{+,f}^{(j)}(x;E), j=1,2,$  coincide for $|x|\geq R_1$, i.e.,

\beq
\phi_{+,f}^{(1)}(x;E)=\phi_{+,f}^{(2)}(x;E), \, {\rm for}\, |x| \geq R_1, f \in \ls.
\label{4.8}
\ene
\end{lemma}

\noindent{\it Proof:} We follow the proof of \cite{we1}. Let $A_j$ be the magnetic potentials (\ref{4.4}-\ref{4.6})
corresponding to $F_j$, and let $Q_j$ be defined as in (\ref{2.2}) with $A_j$ instead of $A$, for $j=1,2$. Denote,
\beq
\psi:= \phi^{(2)}_{+,f}- \phi^{(1)}_{+,f},
\label{4.9}
\ene
\beq
\varphi:= Q_1 \phi^{(1)}_{+,f}-Q_2 \phi^{(2)}_{+,f}.
\label{4.10}
\ene

Then,
\beq
(H_0-E)\psi= \varphi
\label{4.11}
\ene
and furthermore, $\psi \in \mathcal H^{2}_{-s}, \varphi \in L^2_s, 0 < s < (1+\rho)/2.$

As $S_1(E)=S_2(E)$ it follows from (\ref{2.11}) and (\ref{2.13}) that,

\beq
\int_{\ere^n}\, e^{\ds -i E^{1/2}x\cdot \omega }\, \varphi(x)\, dx
=0
\label{4.12}
\ene
 in trace sense. Let us denote by $\hat{\varphi}(\xi)$ the Fourier transform of $\varphi$. Then, $\hat{\varphi}\in
 \mathcal H^{s}$ and $\hat{\varphi}(\xi)=0$ in trace sense on the sphere $|\xi|= E^{1/2}$. Furthermore, denoting by
 $\hat{\psi}$ the Fourier transform of $\psi$, it follows from (\ref{4.11}) that,

$$
(\xi^2 -E) \hat{\psi}= \hat{\varphi}(\xi).
$$
Then, by Theorem 3.5 of \cite{ag}, $\hat{\psi} \in \mathcal H^{s-1}$, what implies that
$\psi \in L^2_{s-1}.$

Moreover, as $V_1=V_2, A_1=A_2$ for $|x| \geq R_1$,
$$
(H_0+Q_1 )\psi = E \psi, \, {\rm for}\, |x| \geq R_1,
$$
and since  $s-1 > - 1/2$, it follows from  Theorem 1.2 of
\cite{au}  that $ \psi(x)=0$ for $|x|$ large enough, and then, by
unique continuation  \cite{ho}, \cite{wo}, $\psi(x)=0,$ for $|x| \geq R_1$, and
(\ref{4.8}) holds.

\bull

In our first uniqueness result we consider the case where the magnetic field is identically zero.

\begin{theorem}
Suppose that $F =0$ and let $V_j,j=1,2$ be electric potentials in $\ere^n,n\geq 3,$ that satisfy,
$V_j \in L^n_{\rm loc}$ and such that for some $ R>0$,
\beq
\left| V_j(x) \right|\leq C (1+|x|)^{-\rho}, \quad \rho >1, \,|x| \geq R > 0, j=1,2.
\label{4.13}
\ene
 Let $S_j(E)$ be the scattering matrices corresponding, respectively, to $V_j, j=1,2$.
Then, if for some $E >0$, $S_1(E)= S_2(E)$ and  $V_1(x)=V_2(x)$ for $|x|\geq R >0$,
the electric potentials coincide everywhere, i.e. $V_1(x)= V_2(x), x \in \ere^n.$
\end{theorem}
\noindent{\it Proof:} Let $B_R$ denote the open the ball of center
zero and radius $R$. Let $\varphi_j\in \mathcal H^2(B_R), j=1,2,$
be any solutions to

\beq \left(H_0+V_j- E\right) \varphi_j=0,
j=1,2.
\label{4.14}
\ene

Then, multiplying the equation for $j=1$
by $\overline{\varphi_2}$ and the complex conjugate of the
equation for $j=2$ by $\varphi_1$ integrating over $B_R$
substracting   the resulting equations and using Green's formula
we obtain the following identity,

\beq
\int_{\ds B_R} \,(V_1-V_2)\, \varphi_1\, \overline{\varphi_2}\, dx= \int_{\ds \partial B_R}\, \left(\overline{\varphi_2}
\partial_{\nu}\varphi_1- \varphi_1 \overline{\partial_{\nu}\varphi_2}\right)\, dS,
\label{4.15}
\ene
where $\nu$ is the exterior unit normal  to $\partial B_R$.

It follows from Lemma 4.1  that

\beq
\int_{B_R}\,\left(V_2-V_1\right) \phi_{+,f}^{(1)}\, \overline{ \phi_{+,g}^{(2)}}\, dx =0, \, f,g \in \ls.
\label{4.16}
\ene

Then, by Theorem 3.1,
\beq
\int_{\ds B_R} \,(V_2-V_1)\, \varphi_1\, \overline{\varphi_2}\, dx=0
\label{4.17}
\ene
for every $\varphi_j \in \mathcal H^2(B_R) $ that are solutions to (\ref{4.14}), j=1,2.

Now as in \cite{we1} for any $ p \in \CE^n, p^2=E, {\mathrm Im} p \neq 0,$ and $|p|$ large enough
 we construct solutions $\varphi_j(x,p)\in \mathbb H^{2}_{\mathrm loc}(\ere^n)$
to the equations
$$
(H_0+ \chi_{\ds B_R}(x)V_j) \varphi_j(x,p) = E \varphi_j(x,p),j=1,2,
$$
where $\chi_{\ds B_R}$ is the characteristic function of $B_R$, such that,

$$
\varphi_j(x,p)= e^{i p \cdot x}(1+ \psi_j(x,p)),
$$
where $ \|\psi_j(x,p)\|_{\ds \mathbb H^1_{-s}}\| \leq C_s, s > 1/2,$ and
$$
{\mathrm s-}\lim_{|p| \to \infty}\|\psi_j(x,p)\|_{\ds L^2_{-s}}=0.
$$
For this purpose, note that as $V \in L^n_{\rm loc}$ and (\ref{4.7}) holds it follows from Theorem 5.2 of
\cite{sch}  that $(1+|x|)^{\rho} V_j, j=1,2$ are bounded
operators from $\mathcal H^1$ into $L^2$ and that their norm can be bounded by a constant times $1+ \|V_j\|_{L^n( B_R)}$.

Given any $\xi \in \ere^n$ take a sequence $p^{(j)}_l$ satisfying $ \left(p^{(j)}_l\right)^2= E,
{\rm Re}\, p^{(1)}_l-{\rm Re}\, p^{(2)}_l= \xi, {\rm Im}\, p_l^{(1)}=
-{\rm Im}\, p^{(2)}_l\neq 0, \lim_{l \to \infty}|p^{(j)}|_l=\infty$. This is possible as $n \geq 3.$
Since $\varphi\left(x,p^{(j)}_l\right)$ are solutions to (\ref{4.14})
in $B_R$ we have that,
$$
\int_{\ds B_R} \,(V_2(x)-V_1(x))\, \varphi_1(x,p^{(1)}_l)\, \overline{\varphi_2(x,p^{(2)}_l)}\, dx=0.
$$
But then,

$$
 \int_{\ds B_R} e^{i\xi\cdot x}\, (V_2(x)-V_1(x))\, dx =
 \lim_{l \to \infty}\,\int_{\ds B_R} \,(V_2(x)-V_1(x))\, \varphi_1(x,p^{(1)}_l)\, \overline{\varphi_2(x,p^{(2)}_l)}\,
 dx = 0,
$$
and it follows that $V_1(x)=V_2(x), x \in B_R$.

\bull

We now consider the case where there is also a magnetic potential.

\begin{theorem}
Suppose that  $ V_j, F_j \in C^{\infty}(\ere^n), n\geq 3$, that
$V_j$ satisfies (\ref{4.7}) and $F_j$ satisfies (\ref{4.3}), j=1,2.
Let $S_j(E)$ be the scattering matrices corresponding,
respectively, to $(V_j,F_j), j=1,2$. Then, if some $E >0$, $S_1(E)=
S_2(E)$ and  $V_1(x)=V_2(x), F_1(x)=F_2(x)$ for $|x|\geq R >0$, we have that the
electric potentials and the magnetic fields coincide everywhere,
i.e. $V_1(x)= V_2(x), F_1(x)=F_2(x), x \in \ere^n.$
\end{theorem}
\noindent{\it Proof:} Let us consider the following Dirichlet
problems on $B_R$,

\beq
(H_0+ Q_j -E )\varphi =0, \, \varphi|_{\ds\partial B_R}=f,
\label{4.18}
\ene
where $Q_j,j=1,2,$ are defined as in (\ref{2.2}) with $V_j, A_j$ instead of $V,A$ and  $A_j$  given by
(\ref{4.4})-(\ref{4.6}) with the corresponding $F_j$. We can always take a $R$ such that zero is not an
eigenvalue of $H_0+Q_j$ for both, $j=1$ and $j=2$
\cite{le}. Then, for every $f\in \mathcal H^{1/2}(\partial \Omega)$ there is a unique $\varphi_j \in \mathcal H^1(B_R)$
that solves (\ref{4.18}) for $j=1,2$. The Dirichlet to Neumann maps, $\Lambda_j,$ are the operators mapping
$\mathcal H^{1/2}(\partial \Omega)$ into $\mathcal H^{-1/2}(\partial \Omega)$ defined as,

\beq
\Lambda_j f= \frac{\partial}{\partial \nu}\varphi_j|_{\ds\partial B_R}-i A_j\cdot\nu f, j=1,2,
\label{4.19}
\ene
where $\nu$ is the exterior unit normal to $\partial B_R$ and with
$\frac{\partial} {\partial\nu}\varphi_j\big|_{\ds\partial B_R}$
defined in trace sense.

 The following identity is proven by integration by
parts \cite{sun},

\beq
\begin{array}{ll}
 i\int_{B_R}\left(A_2-A_1\right)\cdot \left(\varphi_1\, \nabla
\overline{\varphi_2}-\overline{\varphi_2}\,\nabla \varphi_1\right)\, dx
+\int_{B_R}\, (A_1^2-A_2^2+V_1-V_2)\varphi_1 \, \overline{\varphi_2}  \, dx \\ \\
=-\int_{\ds \partial B_R}\, \overline{\varphi_2}\, [\Lambda_1 -\Lambda_2]\varphi_1 \, dS,\\
\end{array}
\label{4.20}
\ene
where $\varphi_j \in \mathcal H^1(B_R)$ are any pair of solutions to $(H_0+ Q_j -E )\varphi_j =0, j=1,2$.

Since by Lemma 4.1, $\phi^{(1)}_{+,f}= \phi^{(2)}_{+,f},\frac{\partial}{\partial \nu}\phi^{(1)}_{+,f}=
\frac{\partial}{\partial \nu}\phi^{(2)}_{+,f} $  and $A_1=A_2$ on $\partial B_R$ we have that,

$$
\Lambda_2 \phi^{(1)}_{+,f}|_{\ds \partial B_R}= \Lambda_2 \phi^{(2)}_{+,f}|_{\ds \partial B_R}=
\frac{\partial}{\partial \nu}\phi^{(2)}_{+,f}|_{\ds \partial B_R}-iA_2 \phi^{(2)}_{+,f}|_{\ds \partial B_R}
= \Lambda_1 \phi^{(1)}_{+,f}|_{\ds \partial B_R},
$$
and taking $\varphi_1= \phi^{(1)}_{+,f}$ in (\ref{4.20}) we obtain that,

\beq
i\int_{B_R}\left(A_2-A_1\right)\cdot \left(\phi^{(1)}_{+,f}\, \nabla
\overline{\varphi_2}-\overline{\varphi_2}\,\nabla \phi^{(1)}_{+,f}\right)\, dx
+\int_{B_R}\, (A_1^2-A_2^2+V_1-V_2)\,\phi^{(1)}_{+,f} \, \overline{\varphi_2}  \, dx =0.
\label{4.21}
\ene

Let $\phi^{(1)}_{+,f_n}$ be a sequence that converges to $\varphi_1$ in $L^2(B_R)$.
Then,

\beq
\begin{array}{ll}
\lim_{n \to \infty} \int_{B_R}\,(A_2-A_1)\cdot\overline{\varphi_2}\,\nabla \phi^{(1)}_{+,f_n}\, dx
= - \lim_{n \to \infty} \int_{B_R}\,\left[\nabla ((A_2-A_1)\cdot\overline{\varphi_2})\right]\phi^{(1)}_{+,f_n}\,
 dx =
\\
\\
- \int_{B_R}\,\left[\nabla((A_2-A_1)\cdot\overline{\varphi_2})\right]\, \varphi_1\, dx
=\int_{\ds B_R}\, (A_2-A_1)\cdot\overline{\varphi_2} \, \nabla \varphi_1\, dx.
\\
\end{array}
\label{4.22}
\ene
Replacing $\phi^{(1)}_{+,f}$ by $\phi^{(1)}_{+,f_n}$ in (\ref{4.21}), taking the limit when $n \to \infty$
and using (\ref{4.22}) we obtain that

$$
i\int_{B_R}\left(A_2-A_1\right)\cdot \left(\varphi_1 \nabla \,
\overline{\varphi_2}-\overline{\varphi_2}\, \nabla \varphi_1\right)\, dx
+\int_{B_R}\, (A_1^2-A_2^2+V_1-V_2)\varphi_1 \, \overline{\varphi_2}  \, dx
= 0.
$$
But as this holds for any $\varphi_j \in \mathcal H^{(1)}(B_R)$ with $(H_0+Q_j -E)\varphi_j=0, j=1,2,$ it follows from
(\ref{4.20}) that $\Lambda_1=\Lambda_2$. Then, by Theorem (B) of \cite{nus},  $V_1=V_2, F_1=F_2$ on $B_R$
and the theorem follows.

\bull

In order to state our uniqueness result for electric potentials and magnetic fields with a regular behaviour at
infinity we introduce some notation \cite{wy}.

Let us  denote by $\dot{{\mathcal S}}^{-\rho}$ the set
of $C^\infty({\mathbb R}^n\setminus\{0\})$-functions $f(x)$ such that
$\partial^\alpha f(x)=O(|x|^{-\rho-|\alpha|})$ as $|x|
\rightarrow\infty$ for all $\alpha$. An
important example of functions from the class  $\dot{{\mathcal
S}}^{-\rho}$ are homogeneous functions $f \in C^\infty({\mathbb
R}^n\setminus\{0\})$ of order $-\rho $ such that $f(\lambda x)
=\lambda^{-\rho} f(x)$ for all $x\in {\mathbb R}^n, x\neq 0,$ and $\lambda  > 0$.

Let the functions $f_{l}\in \dot{{\mathcal S}}^{-\rho_{l}}$   where $\rho_{l}\rightarrow\infty$ (but the condition
 $\rho_{l}<\rho_{l+1}$ is not required). The  notation

\beq
f (x) \simeq \sum_{l=1}^\infty f_l(x)
 \label{4.23}
 \ene
 means that, for any $N$,   the remainder

 \beq
 f   - \sum_{l=1}^N f_j \in\dot{\mathcal S}^{-\rho}\quad \mathrm{where} \quad \rho=\min_{l\geq N+1} \rho_{l} .
  \label{4.24}
  \ene
  In particular, if the sum (\ref{4.23}) consists of a finite number $N$ of terms, then the inclusion (\ref{4.24})
   should be satisfied for all $\rho$.
A function $f \in C^\infty$ is determined by its asymptotic expansion (\ref{4.23}) up to a term from the Schwarz
 class ${\mathcal S}={\mathcal S}^{-\infty}$.

\begin{theorem}
Let the electric potentials $V_j$ and the magnetic fields $F_j$ be $C^{\infty}(\ere^n)$ functions, $ n \geq 3,j=1,2,$
and assume that
they satisfy,

\beq
   \left|\partial^{\alpha} V_j(x) \right|\leq C (1+|x|)^{-\rho-|\alpha|},
   \left|\partial^{\alpha} F_j(x)\right|  \leq C
 (1+|x|)^{- 1-\rho-|\alpha|}, \quad \rho > 1,
\label{4.25}
\ene
for all $\alpha$. Moreover, suppose that they admit the
asymptotic expansions
\beq
V_j (x) \simeq \sum_{l=1}^\infty V_{j,l}(x), \quad  F_j (x) \simeq \sum_{l=1}^\infty F_{j,l}(x), j=1,2,
 \label{4.26}
 \ene
where $V_{j,l}$ and $F_{j,l}$ are homogeneous functions of orders, respectively, $-\rho_{j,l}$ and $-r_{j,l},$
with, $1 < \rho_{j,1} < \rho_{j,2} < \cdots$, and $ 2 < r_{j,1} < r_{j,2} < \cdots, j=1,2. $
Assume, moreover, that the asymptotic expansions (\ref{4.26}) actually converge, respectively, to $V_j$ and
$F_j, j=1,2,$ in pointwise sense, for $|x|$ large enough, or just that the sums in (\ref{4.26}) are  finite.
Let $S_j(E)$ be, respectively, the scattering matrices corresponding to $(V_j,F_j), j=1,2.$
Then, if for some $E >0$, $S_1(E)= S_2(E)$, we have that $V_1(x)=V_2(x)$ and $F_1(x)=F_2(x), x \in \ere^n.$
\end{theorem}
\noindent {\it Proof:} By Theorem 4.2 of \cite{wy} $V_{1,l}= V_{2,l}, F_{1,l}=F_{2,l}, l=1,2,\cdots$. Moreover, since
the asymptotic expansions in (\ref{4.26}) actually converge, respectively, to $V_j$, $F_j$ for
$|x|$ large enough, we have that $V_1(x)=V_2(x), F_1(x)=F_2(x)$ for  $|x|$ large enough, and then by Theorem 4.3
$V_1(x)=V_2(x), F_1(x)=F_2(x), x \in \ere^n$. The same argument applies if the sums in (\ref{4.26}) are finite.


\begin{thebibliography}{99}

\bibitem{ab} Aharonov Y and Bohm D, Significance of electromagnetic potentials in the quantum theory,
Phys. Rev. {\bf 115} (1959), 485-491

 \bibitem{ag} Agmon S, Spectral properties of Schr\"odinger operators and scattering theory, Ann. Sc.
 Norm. Super. Pisa Cl. Sci. {\bf II}, 2 (1975), 151-218

\bibitem{au} Arai M and Uchiyama J, Growth order of eigenfunctions of Schr\"odinger operators with potentials
admitting some integral condition I- general theory, Publ. Res. Inst. Math. Sci.  Kyoto Univ. {\bf 32} (1996), 581-616

\bibitem{cs} Chadan K and Sabatier S, {\it Inverse Problems in
Quantum Scattering Theory, 2nd ed.,} Springer, Berlin, 1989.


\bibitem{ei}  Eidus D,  Completeness properties of scattering problem solutions, Comm.  Partial Differential
Equations, {\bf 7} (1982), 55-75

\bibitem{er} Eskin G and Ralston J, Inverse scattering problem
for the Schr\"odinger equation with magnetic potential at a fixed
energy,  Comm. Math. Phys. {\bf 173} (1995), 199-224


\bibitem{Gr}  Grinevich P G, Rational solitons of the
Veselov-Novikov equations and reflectionless two-dimensional
potentials at fixed energy,  Teoret. Mat. Fiz. {\bf 69} (1986), 307-310
[English transl. in  Theoret. and Math. Phys. {\bf 69} (1986),
1170-1172]

\bibitem{gr} Guillot J C and Ralston J,  Inverse scattering
at a fixed energy for layered media,  J. Math. Pures Appl.
{\bf 78} (1999), 27-48

\bibitem{ho} H\"ormander L, Uniqueness theorems for second order elliptic differential equations, Comm.  Partial
Differential Equations {\bf 8} (1983), 21-64.

\bibitem{is} Isakov V, {\it Inverse Problems for Partial Differential Equations,} Applied Mathematical Sciences
{\bf 127} Springer, Berlin, 1998

\bibitem{iso} Isozaki H,  Inverse scattering theory for Dirac
operators, Ann. Inst. H Poincar\'e {\bf 66} (1997), 237-270

\bibitem{iso2} Isozaki H,  Inverse scattering theory for wave
equations in stratified media,  J. Differential Equations {\bf 138} (1997), 19-54


\bibitem{kur}  Kuroda S T,  Scattering theory for differential operators I
operator theory, II  self-adjoint elliptic
operators,  J. Math. Soc. Japan {\bf 25} (1973), 75-104, 222-234

\bibitem{kur2} Kuroda S T, {\it An introduction to Scattering
Theory,} Lecture Notes Series {\bf 51}, Matematisk Institut, Aarhus
Universitet, 1980

\bibitem{le} Leis R, Zur Monotonie der Eigenwerte sebstadjungierter elliptischer Differentialgleichungen,
Math. Z. {\bf 96} (1967), 26-32

\bibitem{nus} Nakamura G, Uhlmann G and Sun Z, Global
identifiability for an inverse problem for the Schr\"odinger
equation with magnetic field,  Math. Anal. {\bf 303} (1995), 377-388

\bibitem{no} Novikov R G, The inverse scattering problem at
fixed energy for the three dimensional Schr\"odinger equation with
an exponentially decreasing potential,  Comm. Math. Phys.
{\bf 161} (1994), 569-595


\bibitem{ra} Ramm A G, {\it Multidimensional Inverse Scattering Problems,} Pitmann Monographs and
Surveys in Applied Mathematics {\bf 51}, Longman/ Wiley, New York,
1992

\bibitem{rs2} Reed M and Simon B, {\it Methods of Modern Mathematical Physics
II  Fourier Analysis, Self-Adjointness,} Academic Press, New York, 1975



\bibitem{rs} Reed M and Simon B, {\it Methods of Modern Mathematical Physics
III Scattering Theory,} Academic Press, New York, 1979



\bibitem{sch} Schechter M, {\it Spectra of Partial Differential Operators, Second Edition,} Applied Mathematics
and Mechanics {\bf 14}, North Holland, Amsterdam, 1986

\bibitem {skr}  Skriganov M  M, Uniform coordinate and spectral asymptotics for  solutions
of the scattering problem   for the Schr\"odinger equation,  J. Soviet
Math. {\bf 8} (1978), 120-141

\bibitem{sun} Sun Z, An inverse boundary value problem for Schr\"odinger operators with vector potentials,
Trans. Amer. Math. Soc {\bf 338} (1993), 953-969


\bibitem{su} Sylvester J and Uhlmann G, The Dirichlet to Neumann map and its applications, in {\it Inverse Problems
in Partial Differential Equations,} (Arcata, CA 1989). Editors R.Colton and W. Rundell, SIAM, Philadelphia, 1990, pp.
101-139





\bibitem{uh} Uhlmann G and Vasy A, Fixed energy inverse
problem for exponentially decreasing potentials,  Methods Appl.
Anal. {\bf 9} (2002), 239-247



\bibitem{we1}  Weder R,   Global uniqueness at a fixed energy in
multidimensional inverse scattering theory,  Inverse Problems
{\bf 7} (1991), 927-938

\bibitem{we2} Weder R, {\it Spectral and Scattering Theory in
Perturbed Stratified Media,} Applied Mathematical Sciences {\bf
87}, Springer, New York, 1991

\bibitem{we3} Weder R, Multidimensional inverse problems in
perturbed stratified media,  J. Differential Equations {\bf 152} (1999), 191-239

\bibitem{we5} Weder R, The Aharonov-Bohm effect and inverse scattering theory,
Inverse Problems {\bf 18} (2002), 1041-1056

\bibitem{we4} Weder R, Inverse scattering at a fixed quasi-energy
for potentials periodic in time,  Inverse Problems {\bf 20} (2004),
893-917

\bibitem{wy} Weder R and Yafaev D R, On inverse scattering at a fixed energy for potentials with a regular
behaviour at infinity, preprint 2005, ArXiv math-ph/0508020, http://arxiv.org/PS\_cache/math-ph/pdf/0508/0508020.pdf,
mp\_arc 05-270, http://www.ma.utexas.edu/mp\_arc-bin/mpa?yn=05-270. To appear in Inverse Problems

\bibitem{wo} Wolff T H, Recent work on sharp estimates in second order elliptic unique continuation
problems, in {\it Fourier Analysis and Partial Differential Equations}, edits. Garc\'{\i}a-Cuevas J, Hern\'andez F S
and Torrea J L, Studies in Advanced Mathematics, CRC Press, Boca Rat\'on, 1995, pp. 99-128.

\bibitem{yaf2} Yafaev D R, On solutions of the Schr\"odinger equation with radiation condition
at infinity, Advances in Soviet Mathematics {\bf 7} (1991), 179-204


\bibitem{yaf4} Yafaev D R, {\it Mathematical Scattering Theory,} AMS, Providence, 1992

\bibitem{yaf1} Yafaev D R, {\it Scattering Theory: Some Old and New
Problems,} Lecture Notes in Math. {\bf 173}, Springer, Berlin, 2000



\bibitem{yaf3} Yafaev D  R,  Scattering by magnetic fields,  St. Petersburg Math. J. {\bf 17} N 5 (2006),
 and arXiv SP/0501544


\end{thebibliography}
\end{document}